\documentclass[pra,showpacs,twocolumn]{revtex4}
\usepackage{graphicx,color}
\usepackage{amssymb}
\usepackage{rotating}
\usepackage{ dsfont }

\newcommand{\nn}{\nonumber}
\newcommand{\eqref}[1]{Eq.~(\ref{#1})}
\newcommand{\ket}[1]{\left|{#1}\right\rangle}
\newcommand{\bra}[1]{\left\langle{#1}\right|}
\newcommand{\scalar}[2]{\left\langle\left.{#1}\right|{#2}\right\rangle}

\begin{document}

\title{A time dependent Markovian master equation for adiabatic systems and its application to the Cooper pair pumping}
\author{I. Kamleitner$^1$, and A. Shnirman$^{1,2}$}
\affiliation{${}^1$Institut f\"ur Theory der Kondensierten Materie, Karlsruher Institut f\"ur Technologie, 76128 Karlsruhe, Germany}
\affiliation{${}^2$DFG-Center for Functional Nanostructures (CFN), 76128 Karlsruhe, Germany}
\begin{abstract}
For adiabatically and periodically manipulated dissipative quantum 
systems we derive, using Floquet theory, a simple Markovian master 
equation. Contrary to some previous works we explicitly take 
into account the time dependence of the Hamiltonian and, therefore, obtain a master equation
with a time-dependent dissipative part. We illustrate our theory with two examples and 
compare our results with the previously proposed master equations. 
In particular, we consider the problem of Cooper pair pumping and demonstrate 
the inadequacy of the secular (rotating wave) approximation when calculating the 
pumped charge. The secular approximation producing a master equation of the 
Lindblad type approximates well the quantum state (density matrix) of the system, 
while to determine the pumped charge a non-Lindblad master equation 
beyond the rotating wave approximation is necessary.  
\end{abstract}
 \maketitle

\section{INTRODUCTION\label{secint}}

    The hope to develop a quantum information processing machine has made a strong impact in several distinct branches of science, e.g., in theoretical and experimental quantum physics, in computer science, and in electrical and chemical engineering. To succeed, it is essential to achieve high-precision quantum transformations, which so far are hindered by two problems: decoherence (see~\cite{notation} for notation) and gate control. While the former can be solved to some extend by proposing new types of less fragile qubits, or implementing decoherence free subspaces~\cite{pal,lid}, the solution to the latter could be the use of adiabatic transformations~\cite{dua}. Such transformations have the advantage that they can be surprisingly robust to various errors of experimental parameters.

    This robustness of adiabatic gates, however, comes at the price of a significantly longer gate time, therefore giving the environment more time to destroy coherence. For this reason, it is important to understand the interplay of adiabaticity and decoherence, which was studied in several publications in connection to Berry's phase~\cite{eli,gam,sjo,kam,sar}. In these, the authors usually postulated a Markovian master equation of Lindblad form.

    However, microscopic derivations of master equations show that its structure crucially depends on the Hamiltonian of the system. Therefore, the time dependence of the Hamiltonian should be explicitly taken into account when deriving a master equation. This approach was already taken, e.g., in Refs.~\cite{rus,sol,pek} which studied the pumped charge in a Cooper pair sluice, as well as in Ref.~\cite{rou} and Ref.~\cite{cai} which studied hysteresis in molecular magnets and entanglement of two spin molecules, respectively. 
In this paper we are interested both in general properties of decoherence in adiabatic systems as well as 
in the concrete problem of Cooper pair pumping. We also discuss the applicability of the secular approximation and show that it leads to an incorrect pumped charge despite the fact that this approximation is very well justified if one is interested in the quantum state (density matrix) of the system. 

	We wish to point out that there are two distinct notions of adiabaticity. First, in the quantum literature one speaks of adiabatic behavior if a closed system approximately follows the instantaneous eigenstates of the Hamiltonian $H(t)$, i.e.\ $\ket{\psi(t)}\approx e^{i\varphi(t)}\ket{n(t)}$ with $H(t)\ket{n(t)}=E_n(t)\ket{n(t)}$. Second, in the thermodynamical literature adiabaticity means that the system follows the instantaneous thermal equilibrium state $\rho(t)\propto \exp[-H(t)/(k_BT)]$, where $T$ is the temperature of the environment. While both situations require a slow change of the Hamiltonian, in the quantum case the time scale is set by the eigenfrequencies of the Hamiltonian, whereas in the thermodynamic case the energy relaxation time $T_{\rm r}$ (often referred to as $T_1$ in the context of qubits) is the appropriate time scale. A master equation describing an adiabatically steered quantum system which is coupled to a thermal bath should therefore recover the thermodynamical adiabatic behavior in the respective limit.

	If $T_{\rm r}$ is much longer than the period $\mathcal T$ of a cyclically time--dependent Hamiltonian, then one would expect that in the quantum adiabatic limit, the system approximately follows a mixture of the instantaneous eigenstates $\rho(t)\approx\sum_m c_m(t)\ket{m(t)}\!\bra{m(t)}$ (or more precisely of Floquet states~\cite{rus}). After a transient period the coefficients $c_n(t)$ approach constant values $c_n(\infty)$ 
which depend on some sort of time--averaged level splittings of the Hamiltonian as well as on the temperature $T$ of the environment. This is the limit which was  described in Ref.~\cite{rus}.

	However, in the other limit, $T_{\rm r}\ll \mathcal T$, one expects that after some transient time the system follows the instantaneous thermal state. The coefficients then satisfy $c_m(t)=c_n(t)\exp[(E_n(t)-E_m(t)/(k_BT)]$ and hence cyclically depend on time. Our work represents an extension of Ref.~\cite{rus} as both limits as well as the crossover between them can be described by our master equations~\eqref{Lindblad} and \eqref{nonLindblad}.

	In section~\ref{secflo} we introduce the Floquet formalism for cyclic Hamiltonians. We then modify the formalism to separate the fast dynamics which arise from the energy splitting of the Hamiltonian from the slow dynamics which are due to the adiabatic time dependence of the Hamiltonian. This procedure directly points to an approximation suitable for adiabatically driven, Markovian open quantum systems, which results in a Lindblad type master equation, but with time--dependent Lindblad operators. We find that decoherence is best described in the basis of the time--dependent Floquet states in the sense that  in this basis one can differentiate between dephasing (the environmental influence on the off--diagonal elements of the density operator) and relaxation (diagonal elements). The decoherence rates generally depend on time to guarantee that in the thermodynamical adiabatic limit the system follows the instantaneous thermal state. It follows that a system coupled to a zero temperature environment does not relax into the instantaneous ground state of the Hamiltonian, but into the slightly different Floquet ground state. We apply our theory in two simple examples. In one of these examples, the problem of 
Cooper pair pumping, the measurable observable is the transferred (pumped) charge. We show that although the Lindblad master equation leads to a very good approximation for the density matrix of the system, it predicts a completely wrong transferred charge. To rectify the problem we refrain from the secular approximation and 
obtain a suitable master equation which is not of Lindblad form.

\section{FLOQUET THEORY\label{secflo}}

	In this section, we first present some well--known results of the Floquet theory~\cite{gri,gue}. We then modify this theory to allow for a separation of fast oscillations due to the energy of the states from the slow oscillations due to the periodic nature of the Hamiltonian.

	For a time--periodic Hamiltonian $H(t)=H(t+\mathcal T)$ there exist solutions of the Schr\"odinger equation of the form
	\begin{eqnarray}
		\ket{\Psi_\alpha(t)} &=& e^{-i\epsilon_\alpha t} \ket{\Phi_\alpha(t)}, \label1
	\end{eqnarray}
	where the Floquet modes $\ket{\Phi_\alpha(t)}=\ket{\Phi_\alpha(t+\mathcal T)}$ have the same period as the Hamiltonian, and $\epsilon_\alpha$ are the quasi energies. The quasi energies are only uniquely defined up to multiples of $\hbar\Omega$ with $\Omega=2\pi/\mathcal T$, because $e^{-in\Omega t}\ket{\Phi_\alpha(t)}$ for all $n\in\mathds{Z}$ correspond to the same Floquet mode but with a shifted energy.

\paragraph*{\bf Adiabatic Floquet theory.}

	For our purposes, it is advantageous to separate the fast dynamics associated with the instantaneous eigenfrequencies of the Hamiltonian from the slow dynamics associated with the time dependence of $H(t)$. Such a separation is not provided by \eqref{1} because the phase factors in \eqref{1} oscillate with constant frequencies $\epsilon_\alpha$. Therefore, if the eigenfrequencies depend on time, as is generally the case, the Floquet modes $\ket{\Phi_\alpha(t)}$ also oscillate rapidly. A more convenient representation is provided by the states $\ket{\phi_\alpha(t)}$, which differ from the Floquet modes only by a time dependent phase factor:
	\begin{eqnarray}
		\ket{\Psi_\alpha(t)} &=& e^{-i\int_0^tdt'\, \bar E_\alpha(t') } e^{-i\bar\varphi_\alpha t/\mathcal T} \ket{\phi_\alpha(t)}. \label3
	\end{eqnarray}
	Here, $\bar E_\alpha(t)=\bra{\phi_\alpha(t)}H(t)\ket{\phi_\alpha(t)}$ is the instantaneous energy of the Floquet mode, and $\bar\varphi_\alpha$ is chosen such that the states $\ket{\phi_\alpha(t)}=\ket{\phi_\alpha(t+\mathcal T)}$ are periodic in time. In fact, it can be easily shown that $\bar\varphi_\alpha=i\int_0^{\mathcal T} dt \bra{\phi_\alpha(t)}\frac{d}{dt}\ket{\phi_\alpha(t)}$ is the non-adiabatic geometric phase (Aharonov--Anandan phase) \cite{aha} associated with the state $\ket{\phi_\alpha(t)}$.

	Note that the ambiguity in the quasi energies of the original Floquet states is non--existent in the modified Floquet states. Therefore, we can now speak of a well defined Floquet ground state and Floquet excited states. Although the concept of a Floquet ground state is not very meaningful if the driving is fast, for adiabatic driving it will turn out to have the same importance as the energy ground state for non--driven systems. Indeed, the Floquet ground state approaches the instantaneous ground state in limit of $\mathcal T\to\infty$.

	In the adiabatic limit, a system prepared in an energy eigenstate will stay in the respective instantaneous eigenstate. Because the eigenstates are cyclic, they are identical to the modified Floquet states in a zeroth order adiabatic approximation. Therefore, if the time evolution is adiabatic, the following approximations are valid
	\begin{eqnarray}
		\ket{\phi_\alpha(t)} &=& \ket{n_\alpha(t)} + \mathcal O(\mathcal A) , \nn\\
		\bar E_\alpha(t) &=& E_\alpha(t) + \mathcal O(\mathcal A^2) , \nn\\
		\bar\varphi_\alpha &=& \varphi_\alpha + \mathcal O(\mathcal A) .  \label4
	\end{eqnarray}
	Here, $\ket{n_\alpha(t)}$ and $E_\alpha(t)$ are the instantaneous eigenstates and energies of the Hamiltonian, respectively, $\varphi_\alpha$ is the adiabatic geometric phase (Berry phase), and  $\mathcal A=\max_{t,\alpha,\beta}\frac{\hbar|\bra{n_\beta(t)}\frac{d}{dt}\ket{n_\alpha(t)}|}{|E_\beta-E_\alpha|}$ is the adiabatic parameter. These modified Floquet modes have the useful property to vary only slowly in time, such that the expansion
	\begin{eqnarray}
		\ket{\phi_\alpha(t)} &=& \sum_{k=-\infty}^\infty \ket{c_{\alpha,k}} e^{-ik\Omega t} \label5
	\end{eqnarray}
	can be restricted to $|k|\Omega\ll\min_{\alpha,\beta,t}|E_\alpha(t)-E_\beta(t)|$. This property will turn out to be useful in the derivation of a master equation.

\paragraph*{\bf Superadiabatic expansion.}
	 Later on we will find that the system decoheres into the Floquet states. That is why we want to have a closer look at the relation of the Floquet states to the instantaneous energy eigenstates. The statement that an adiabatically evolving system follows the instantaneous eigenstates of the Hamiltonian $H(t)$ is only correct to zeroth order of the adiabatic parameter. It is well known~\cite{ber} that if the Hamiltonian is an analytic function in time, then the system follows more closely the eigenstates of the operator
	\begin{eqnarray}
		\widetilde H &=& H + iU^\dag_0\dot U_0 + iU^\dag_0 U^\dag_1\dot U_1 U_0 +\cdots,  \label{superad}
	\end{eqnarray}
	where we omitted the time dependence of the operators for shorter notation and used a dot to indicate the time derivative. The operator $U_0(t)$ is defined by $\ket{n(t)}=U^\dag_0(t) \ket{n(0)}$. Similarly, $U_1(t)$ is defined by $\ket{n_1(t)}=U^\dag_1(t) \ket{n_1(0)}$, where $\ket{n_1(t)}$ are the eigenvectors of the transformed Hamiltonian $H_1(t)=U_0(t)H(t)U_0^\dag(t)+i\dot U_0(t)U_0^\dag(t)$. The first term in \eqref{superad} is of zeroth order in the adiabatic parameter, the second of first order, and so on.

	One should note that for any given adiabatic parameter, the series \eqref{superad} does not (except for some special cases) converge~\cite{ber}. The smaller the adiabatic parameter, the longer the series improves before it starts to diverge. If the series does converge, then the system will exactly follow the eigenstates of $\widetilde H$ which are then the exact Floquet states. Even if the series does not converge, in an approximation which improves exponentially with the adiabatic parameter, one can use the first few summands of \eqref{superad} to obtain the Floquet states. Thus, by using the Floquet modes our theory implicitely takes into account terms of higher order in the adiabatic parameter.

\section{Master Equation\label{sec3}}

	In this section, we derive a master equation in the modified Floquet basis. Some properties of this basis will lead to approximations which are appropriate for adiabatically evolving systems.

	We assume that the total system evolves under the influence of the Hamiltonian
	\begin{eqnarray}
		H_{AB}(t) &=& H_A(t) + H_B + A\otimes B, \label6
	\end{eqnarray}
	where $H_A$ and $H_B$ are the Hamiltonians of the system of interest and the environment, respectively, and $A\otimes B$ is their coupling with $A$ acting on the Hilbert space of the system of interest and $B$ acting on the environments Hilbert space. We assume an initially uncorrelated state $\rho(0)=\rho_A(0)\otimes\rho_B$ where $\rho_B$ describes the thermal equilibrium of the environment. Without loss of generality we assume Tr$(\rho_BB)=0$, and using the Born-Markov approximation we arrive at the well known interaction picture master equation~\cite{bre}
	\begin{eqnarray}
		\dot\rho_A^I(t) &=& \int_0^\infty d\tau\, {\rm{Tr}}_B\!\left[B(\tau)B\rho_B\right] \label{7}\\
		&&\times \left[ A(t\!-\!\tau)\rho_A(t)A(t)-A(t)A(t\!-\!\tau)\rho_A(t) \right] +c.c., \nn
	\end{eqnarray}
	where $A(t)$ and $B(t)$ denote operators in the interaction picture. We note that the unitary evolution operator used to transform $A$ into the interaction picture may be written conveniently as
	\begin{eqnarray}
		U_A(t) = \sum_\alpha \ket{\Psi_\alpha(t)}\!\bra{\phi_\alpha}, \label{8}
	\end{eqnarray}
	where we introduce the notation $\ket{\phi_\alpha}=\ket{\phi_\alpha(0)}$. We use this form of the evolution operator to write
	\begin{eqnarray}
		A(t) &=& U_A^\dag(t)AU_A(t) \nn\\
		 &=& \sum_{\alpha\alpha'} \bra{\Psi_{\alpha}(t)}A\ket{\Psi_{\alpha'}(t)} \ket{\phi_{\alpha}}\!\bra{\phi_{\alpha'}}  \nn\\
		&=& \sum_{\alpha\alpha' k} e^{-i\!\int_0^t dt'\,[\omega_{\alpha\alpha'}(t') -k\Omega]}A_{\alpha\alpha',k} \ket{\phi_{\alpha}}\!\bra{\phi_{\alpha'}}, \nn\\ \label{9}
	\end{eqnarray}
	where we used the Fourier expansion
	\begin{eqnarray}
		A_{\alpha\alpha',k} &=& \frac1T\int_0^{\mathcal T} dt\, e^{i\Omega kt} \bra{\phi_{\alpha}(t)}A\ket{\phi_{\alpha'}(t)} , \quad \label{10}
	\end{eqnarray}
	as well as
	\begin{eqnarray}
		\omega_{\alpha\alpha'}(t) &=& \bar E_{\alpha'}(t)-\bar E_{\alpha}(t) + ( \bar\varphi_{\alpha'}-\bar\varphi_{\alpha})/\mathcal T  \quad\label{11}
	\end{eqnarray}
	For most purposes, the angular frequencies $\omega_{\alpha\alpha'} = (E_{\alpha'}-E_\alpha)[1+\mathcal O(\mathcal A)]$ may be approximated by the instantaneous transition frequencies.

	We now substitute \eqref{9} into \eqref{7} to find the Bloch-Redfield type equation
	\begin{eqnarray}
		\dot\rho_A^I(t) &\!=\!& \!\sum_{\alpha\alpha'\beta\beta' kl}\!\!  \Gamma[\omega_{\alpha\alpha'}(t)\!+\!k\Omega] A_{\alpha\alpha',k}A_{\beta'\beta,l} \nn\\
		&\!\times\!& e^{i\!\int_0^t dt'[\omega_{\beta\beta'}(t')-\omega_{\alpha\alpha'}(t') - (k+l)\Omega]} \nn\\
		&\!\times\!& \left[ \ket{\phi_\alpha}\!\bra{\phi_{\alpha'}} \rho_A(t)\ket{\phi_{\beta'}}\!\bra{\phi_{\beta}} \right.\nn\\
		&& \left. -\ket{\phi_{\beta'}}\!\scalar{\phi_\beta}{\phi_\alpha}\!\bra{\phi_{\alpha'}}\rho_A(t) \right] \nn\\
		&\!+\!&  \rm{c.c.}, \label{12}
	\end{eqnarray}
	where we define the one-sided Fourier (Laplace) transform of the environment correlation function
	\begin{eqnarray}
		\Gamma[\omega_{\alpha\alpha'}(t)\!+\!k\Omega] &\!\!=\!\!& \int_0^\infty\!\! d\tau\, e^{i\!\int_{t-\tau}^t\!\!\! dt' [\omega_{\!\alpha\alpha'}(t')+k\Omega]} {\rm{Tr}}_B\![B(\tau)B\rho_B] \nn\\
		&\!\!\approx\!\!& \int_0^\infty\!\! d\tau\, e^{i [\omega_{\!\alpha\alpha'}(t)+k\Omega]\tau} {\rm{Tr}}_B\![B(\tau)B\rho_B].
	\end{eqnarray}
	The latter form, which is more familiar from the theory of time independent systems, is valid if the eigenenergies do not change much on the time scale set by the correlation time of the environment.
	
	Using the standard Floquet modes, the master equation~(\ref{12}) was also derived in the review of Grifoni and H\"anggi~\cite{gri}, and served as the starting equation in~\cite{rus}, where the authors proceeded by performing two rotating wave approximations (RWA). The first RWA is to neglect terms with $k\neq l$, and the second one is to keep only terms with $\omega_{\alpha\alpha'}=\omega_{\beta\beta'}$. The first RWA is valid, if $e^{-i(k+l)\Omega t}$ can be considered as fast rotating. In particular for adiabatically evolving systems this might not always be the case, and therefore, we do not use this RWA in the following. 

	We perform an approximation which is particularly suitable for adiabatic driving, namely $\Gamma(\omega+k\Omega)\approx\Gamma(\omega)$. The requirement of a sufficiently smooth correlation function $\Gamma(\omega)$ is satisfied if the modified Floquet modes change slowly compared to the environment correlation time. 

	To arrive at a master equation of Lindblad form and to guarantee complete positivity, we have to perform the second above described RWA, usually referred to as the secular approximation (although, as we will see later, there are some instances where the secular approximation leads to erroneous results and one has to forego complete positivity {\it at all times}). Noting that in general a RWA is valid if the exponential in the neglected term is rotating fast, we recognize from \eqref{12} that this RWA requires that $(k-l)\Omega$ is not of the same order as $\omega_{\alpha\alpha'}-\omega_{\beta\beta'}$. This can only be guaranteed if the Hamiltonian is varied adiabatically, and if the expansion in \eqref{5} can indeed be restricted to only small values of $k$. Therefore, the use of the modified Floquet modes introduced in this paper is essential for this RWA to be valid, even in the adiabatic regime assumed here.

	Noting that $A_{\alpha\alpha',k}$ vanishes unless $k\Omega$ is much smaller than the transition frequency $\omega_{\alpha\alpha'}(t)$, we proceed by neglecting all terms of \eqref{12} with $\omega_{\alpha\alpha'}(t)\neq\omega_{\beta\beta'}(t)$. Using $\sum_k e^{-ik\Omega t} A_{\alpha\beta,k}=\bra{\phi_\alpha(t)}A\ket{\phi_\beta(t)}$ and transforming back into the Schr\"odinger picture, \eqref{12} becomes

	\begin{eqnarray}
		\dot\rho_A(t) &\!=\!& -i[H(t)+H_{\rm{LS}}(t),\rho_A(t)] \nn\\
		&\!+\!& \gamma(0)\left[  L_{0}(t)\rho_A(t)L^\dag_{0}(t)-\frac12 \left\{L^\dag_{0}(t)L_{0}(t),\rho_A(t)\right\}  \right]  \nn\\
		&\!+\!& \sum_{\alpha\neq\beta} \gamma[\omega_{\alpha\beta}(t)] \nn\\
		&\!\times\!& \left[  L_{\alpha\beta}(t)\rho_A(t)L^\dag_{\alpha\beta}(t)-\frac12 \left\{ L^\dag_{\alpha\beta}(t)L_{\alpha\beta}(t),\rho_A(t)\right\}  \right]\!. \nn\\ \label{Lindblad}
	\end{eqnarray}
	with the Lindblad operators
	\begin{eqnarray}
		L_{0}(t) &=& \sum_\alpha \bra{\phi_\alpha(t)}A\ket{\phi_\alpha(t)} \ket{\phi_\alpha(t)}\!\bra{\phi_\alpha(t)},  \qquad \\
		L_{\alpha\beta}(t) &=& \bra{\phi_\alpha(t)}A\ket{\phi_\beta(t)}  \ket{\phi_\alpha(t)}\!\bra{\phi_\beta(t)}. \label{L}
	\end{eqnarray}
	While the operator $L_0$ causes pure dephasing (i.e.\ it does not change the populations of the Floquet modes), the operators $L_{\alpha\beta}$ are responsible for energy relaxation. We further separated the imaginary and real part of the Laplace transformed environment correlation function $\Gamma(\omega)=\frac12\gamma(\omega)+iR(\omega)$ and found the Lamb shift contribution to the Hamiltonian $H_{\rm{LS}}(t)=\sum_{\alpha\beta}R[\omega_{\alpha\beta}(t)] |\!\bra{\phi_\alpha(t)}A\ket{\phi_\beta(t)}\!|^2 \ket{\phi_\beta(t)}\!\bra{\phi_\beta(t)} $ to be diagonal in the Floquet basis.

	Thus, taking into account the explicit time dependence of the Hamiltonian, we showed that the system does not decohere into the instantaneous eigenstates, but into the Floquet states, a result previously obtained by~\cite{rus}. The two bases are equal in the adiabatic limit, but they differ for finite periods $\mathcal T$ as according to \eqref{superad} the Floquet states implicitly include higher order adiabatic contributions. While~\cite{rus} uses rates which are averaged over one period, we take into account the rates $\gamma[\omega_{\alpha\beta}(t)]|\!\bra{\phi_\alpha(t)}A\ket{\phi_\beta(t)}\!|^2$ at each instant of time, which might be important in particular for adiabatic driving. On the other hand, the time dependent rates were already obtained in \cite{cai,rou}, where the authors assumed an infinite period $\mathcal T\to\infty$ and hence found that the system relaxes into a mixture of instantaneous energy eigenstates. Therefore, our theory generalizes previous works and reduces to the known results in the respective limits.
	
	Because of the time dependent rates of \eqref{Lindblad}, we can not expect to find a steady state in the Floquet basis. Nevertheless, there is a quasi stationary state $\rho_A(t+\mathcal T)=\rho_A(t)$ with vanishing off-diagonals in the Floquet basis. If the temperature of the environment is zero, i.e.\ $\Gamma(\omega<0)=0$, then the system evolves into the Floquet ground state.

	In the application of our theory to Cooper pair pumping, we are advised by the well known fact~\cite{pek,sol,sal} that the use of the secular approximation may lead to an error in the transferred charge, despite the fact that it leads to a very good approximation of the actual density operator (as explained in subsection~\ref{cpp}). Therefore, we provide here the Schr\"odinger picture master equation obtained without the secular approximation. This master equation follows from \eqref{12} with $\Gamma(\omega+k\Omega)\approx\Gamma(\omega)$. Thus we obtain
	\begin{eqnarray}
	\label{nonLindblad}
		\dot\rho_A(t) &\!=\!& -i[H_A(t),\rho_A(t)] + K(t)\rho_A(t)A^\dag + A\rho_A(t)K^\dag(t) \nn\\
		&& - A^\dag K(t)\rho_A(t) - \rho_A(t)K^\dag(t) A \label{14A}
	\end{eqnarray}
	with
	\begin{eqnarray}
		K(t) = \sum_{\alpha\alpha'} \Gamma[\omega_{\alpha\alpha'}(t)] \bra{\phi_\alpha(t)}A\ket{\phi_{\alpha'}(t)} \ket{\phi_{\alpha}(t)}\!\bra{\phi_{\alpha'}(t)}. \nn\\
	\end{eqnarray}
	This master equation includes the cross-coupling between diagonals and off-diagonals of the density operator. In what follows we will use both, \eqref{Lindblad} and \eqref{14A}, depending on the situation.

\section{Illustrative Examples}

	In this section we illustrate our theory with some simple examples of two-level systems, which could potentially be tested in experiments. In the first, the Floquet states can be found analytically. It serves to demonstrate the time dependence of the energy relaxation and dephasing rates, as well as the relaxation into the Floquet ground state at zero temperature. The second example is slightly more complex as it has a time dependent level--splitting. Therefore, it exhibits a quasi stationary state where the populations in the Floquet basis depend on time.

\subsection{Two level system with constant level--splitting}

	We consider the Hamiltonian 
	\begin{eqnarray}
		H(t) &=& \frac12 B_{\|}\sigma_z + \frac12 B_\bot [\cos(\Omega t)\sigma_x+\sin(\Omega t)\sigma_y] \nn\\
		&\hat=& \frac12 \left(\begin{array}{cc} B_\| & B_\bot e^{-i\Omega t} \\ B_\bot e^{i\Omega t} & -B_\|  \end{array}\right)  ,
	\end{eqnarray}
	where the matrix form is written in the basis $\ket\uparrow,\,\ket\downarrow$ of eigenvectors of $\sigma_z$. The adiabatic parameter is easily obtained to be $\mathcal A=\Omega B_\bot/(B_\bot^2+B_\|^2)$. This Hamiltonian describes, for example, a spin in a magnetic field which rotates around the z-axis. Because the eigenvalues of the Hamiltonian do not depend on time, the modified Floquet states equal the traditional Floquet states:
	\begin{eqnarray}
		\ket{\phi_+(t)} &\!\!=\!\!&  a \ket\uparrow + b e^{i\Omega t} \!\ket\downarrow\!, \nn \\
		\ket{\phi_-(t)} &\!\!=\!\!&  - b \!\ket\uparrow + a e^{i\Omega t}\!\ket\downarrow \! , \label{fl}
	\end{eqnarray}
	with real $a$ and $b$ satisfying
	\begin{eqnarray}
		a/b &=& \frac1{2B_\bot}\!\left[ B_\| -\Omega+\sqrt{(B_\|-\Omega)^2+B_\bot^2} \right] \quad
	\end{eqnarray}
	and $a^2+b^2=1$. The corresponding Floquet energies are given by
	\begin{eqnarray}
		\epsilon_\pm &=& \frac12\left[\Omega \pm \sqrt{(B_\|-\Omega)^2+B_\bot^2}\right],
	\end{eqnarray}
	and are equal to $E_\pm+\varphi_\pm/\mathcal T$.

	\begin{figure}[t]
		\includegraphics[width=0.7\linewidth]{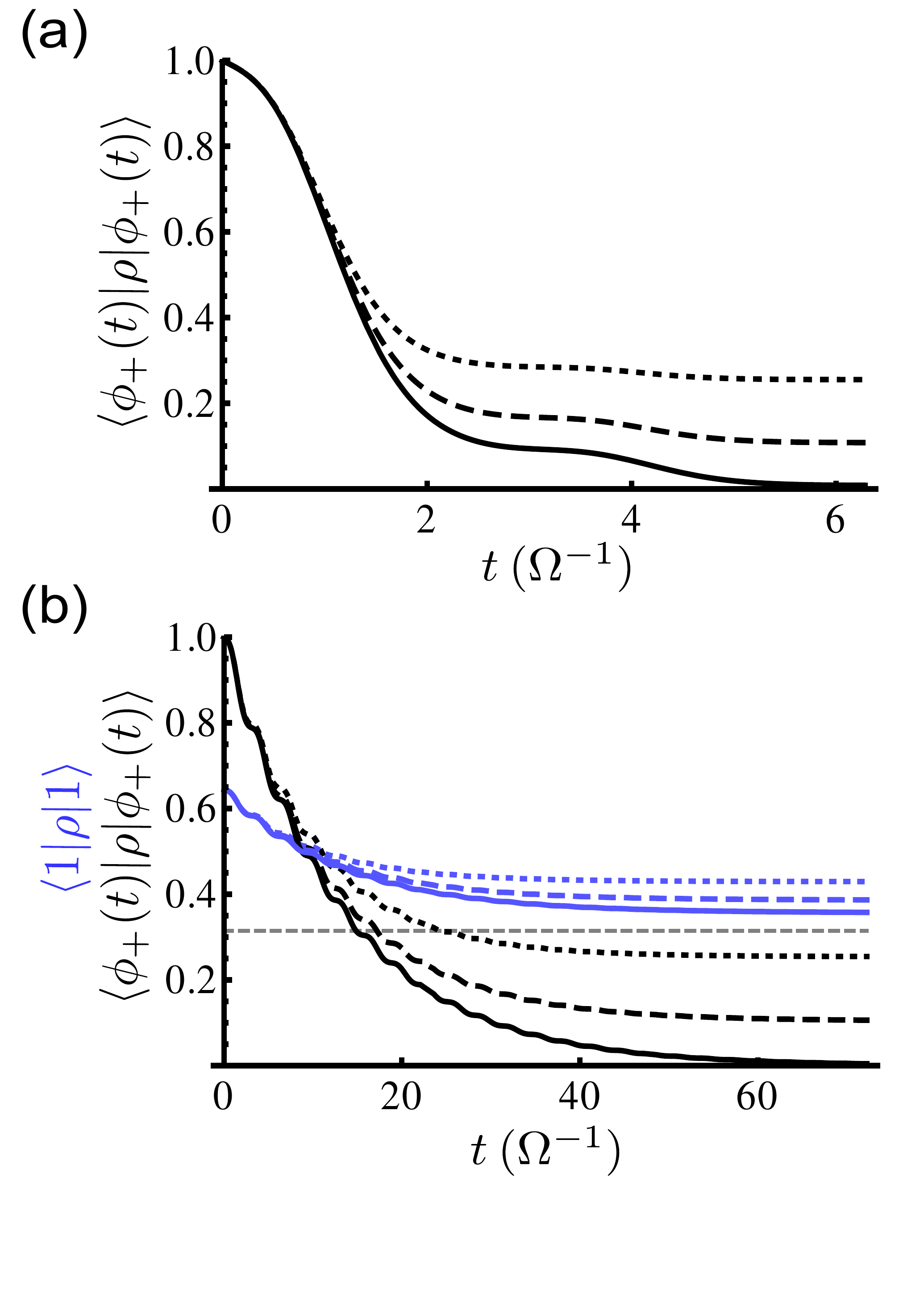}\vspace{-8mm}
		\caption{(Color online) Panel (a) shows the population of the Floquet excited state $\ket{\phi_+(t)}$ during one adiabatic cycle. The damping $\gamma[\omega=(\epsilon_+-\epsilon_-)/\hbar]=0.7~\Omega\hbar^2$ is strong compared to the period $\mathcal T$, such that the quasi stationary state is closely approached within one period. In panel (b) the damping $\gamma[\omega=(\epsilon_+-\epsilon_-)/\hbar]=0.07~\Omega\hbar^2$ is weak such that the quasi stationary state is only reached after many periods. While black lines shows the populations of the Floquet excited state $\ket{\phi_+(t)}$, blue lines are the populations of the state $\ket{\uparrow}$. The gray dashed line is the population of $\ket\uparrow$ if the system would be in the instantaneous ground state. Parameters are chosen as $B_\|/\hbar=2~\Omega$, $B_\bot/\hbar=5~\Omega$, and temperature $k_BT/(2\pi\hbar)=0~\Omega$~(solid), 5~$\Omega$~(dashed), 10~$\Omega$~(dotted). The adiabatic parameter is $\mathcal A=0.043$. \label{fig1}}
	\end{figure}

	We assume a coupling of the system to the environment via $\sigma_x$, which could be a fluctuating external magnetic field in x-direction. Therefore, we find the Lindblad operators in the Schr\"odinger picture from \eqref{L} with $A=\sigma_x$
	\begin{eqnarray}
		L_0 &=&  2ab \cos(\Omega t) \left( \ket{\phi_+(t)}\!\bra{\phi_+(t)} - \ket{\phi_-(t)}\!\bra{\phi_-(t)} \right) , \nn \\
		L_{-+} &=& L_{+-}^\dag = \left( a^2 e^{-i\Omega t} - b^2e^{i\Omega t} \right) \ket{\phi_-(t)}\!\bra{\phi_+(t)}\ , 
	\end{eqnarray}
	We see that there are times when dephasing is the dominant decoherence effect ($L_0$), and times when relaxation ($L_{-+}$) is dominant.

	In Fig.~\ref{fig1}, we show results of the numerical integration of the master equation for an initial state $\ket{\phi_+}$. In panel~(a), the coupling to the environment is strong enough to force a thermal mixture of the Floquet ground and excited states within one period~$\mathcal T$, while in panel~(b), several periods are needed for the system to evolve into the thermal mixture. The populations of the Floquet states approach a constant in the long time limit because the level splitting of the Hamiltonian does not depend on time. The time dependence of the relaxation rate is noticeable in both panels. 

	At zero temperature, the system relaxes into the Floquet--ground state $\ket{\phi_-(t)}$, rather than into the instantaneous ground state of the Hamiltonian. Note that if the time dependence of the Hamiltonian were not explicitly taken into account in the derivation of the master equation, one would find that the system relaxes into the instantaneous ground state of the Hamiltonian. Because the two states have slightly different populations of the state $\ket1$ (the solid blue line does not approach the gray dashed line in the limit $t\to\infty$), the importance of taking into account the time dependence of the Hamiltonian can be tested experimentally by measuring $\sigma_z$ after a sufficiently long time. This effect scales linearly with the adiabatic parameter (see section~\ref{secflo}) and is particularly pronounced if one is in the slightly non-adiabatic regime, which is the case in Fig.~\ref{fig1}.

\subsection{Cooper--pair pumping\label{cpp}}

	\begin{figure}[t]
		\includegraphics[width=\linewidth]{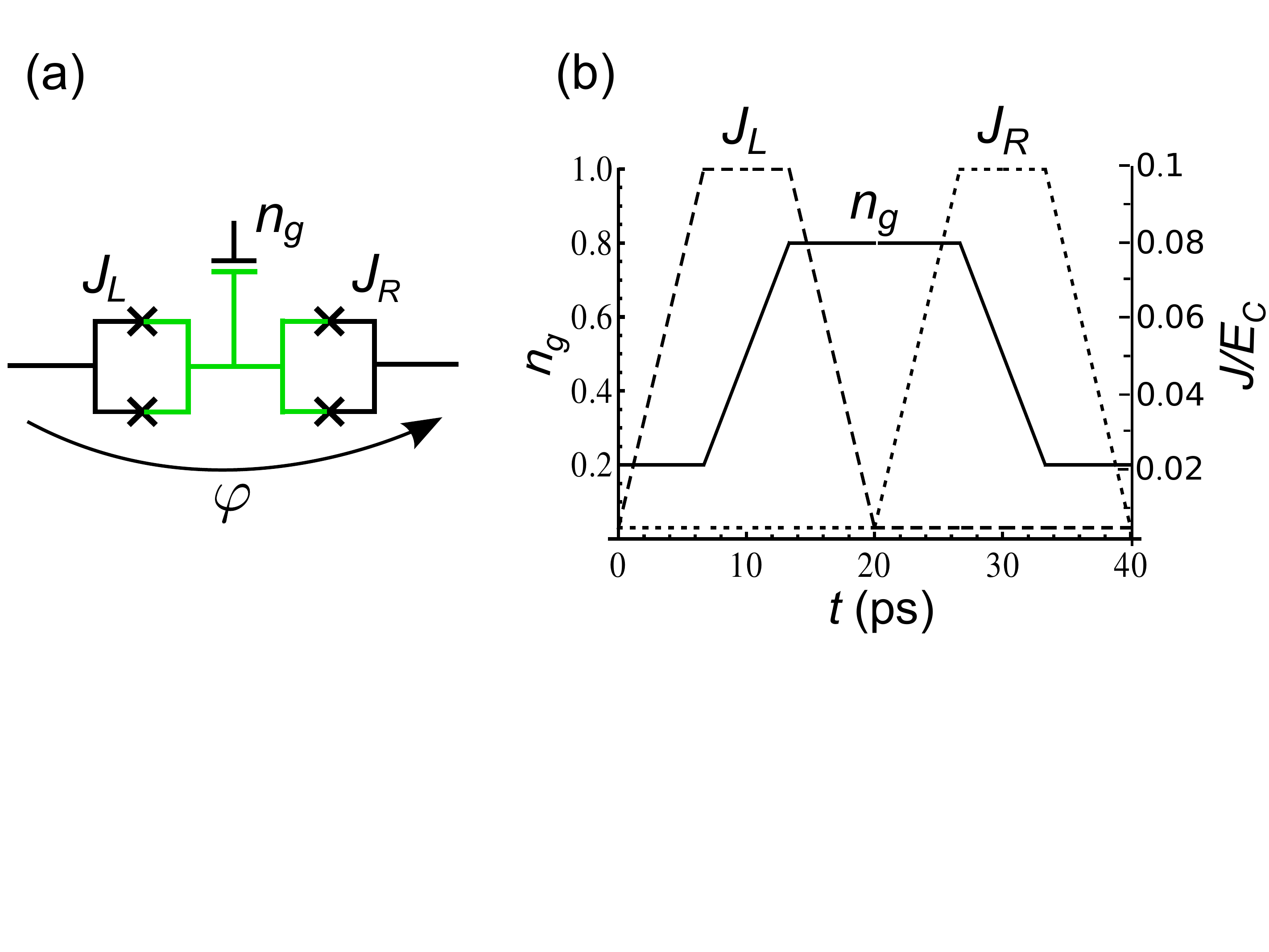} \vspace{-26mm}
		\caption{(Color online) (a) Schematic circuit of a Cooper pair sluice with a superconducting island (green) coupled to a superconducting circuit by two SQUIDs. The electrostatic potential of the island is controlled via a gate voltage.  (b) The pulse sequences for the tunnel rates and the electrostatic potential. Throughout the paper, we use $E_C/(2\pi\hbar)=21$~GHz. \label{fig2}}
	\end{figure}
	In a Cooper pair sluice~\cite{nis,mot2,mot} [see Fig.~\ref{fig2}~(a) for the design], the pumping of Cooper pairs is achieved by a periodic and adiabatic modulation of three system parameters. The effective Josephson couplings $J_{L,R}$ can be controlled by magnetic fluxes through the left and right SQUID, respectively, while the charging energy $E_C(\hat n-n_g)^2$ is controlled via $n_g$ which is tuned by a gate voltage. Here, $E_C=2e^2/C$, where $C$ is the total capacitance of the island, and $\hat n$ is the number operator of the excess Cooper pairs on the superconducting island. Furthermore, the pumped charge depends on the phase difference $\varphi$ of the two leads, which is fixed by embedding the design into a superconducting loop threaded by a fixed magnetic flux. In the following we only consider $\varphi=\pi/2$.

	If the Josephson couplings are small compared to the charging energy, then a two level approximation is valid, where only the two eigenstates of the number operator with the lowest charging energies are considered. The Hamiltonian is then~\cite{mot2}
\begin{eqnarray}
	H &=& -\frac12 (B_x\sigma_x+B_y\sigma_y+B_z\sigma_z), \label{22}\\
	B_x &=& J_L + J_R\cos\varphi,  \nn\\
	B_y &=& J_R\sin\varphi, \nn\\
	B_z &=& E_C(1-2n_g). \nn
\end{eqnarray}

	The pumping pulses are shown in Fig.~\ref{fig2}~(b). Essentially, $n_g$ is raised while the left Josephson junction is open, therefore the island gains one Cooper pair from the left lead. When $n_g$ is lowered, the right junction is open and a Cooper pair is pushed into the right lead. We use a sufficiently slow pumping rate to assure good adiabaticity~\cite{footnote} and without coupling to an environment, the system stays in the Floquet ground state if it was initially prepared in this state. The resulting current ideally is one Cooper pair per pumping cycle. 

	In an experiment, the Josephson junctions can not be closed exactly, and we assume a realistic ratio of $J_{\rm{min}}/J_{\rm{max}}=0.03$. Therefore, in addition to the pumped current which depends on the pumping frequency, there will be a supercurrent which is independent of the pumping frequency~\cite{pek2}. This different frequency dependence can be used to extract the pumped charge.

	\begin{figure}[t]
		\includegraphics[width=\linewidth]{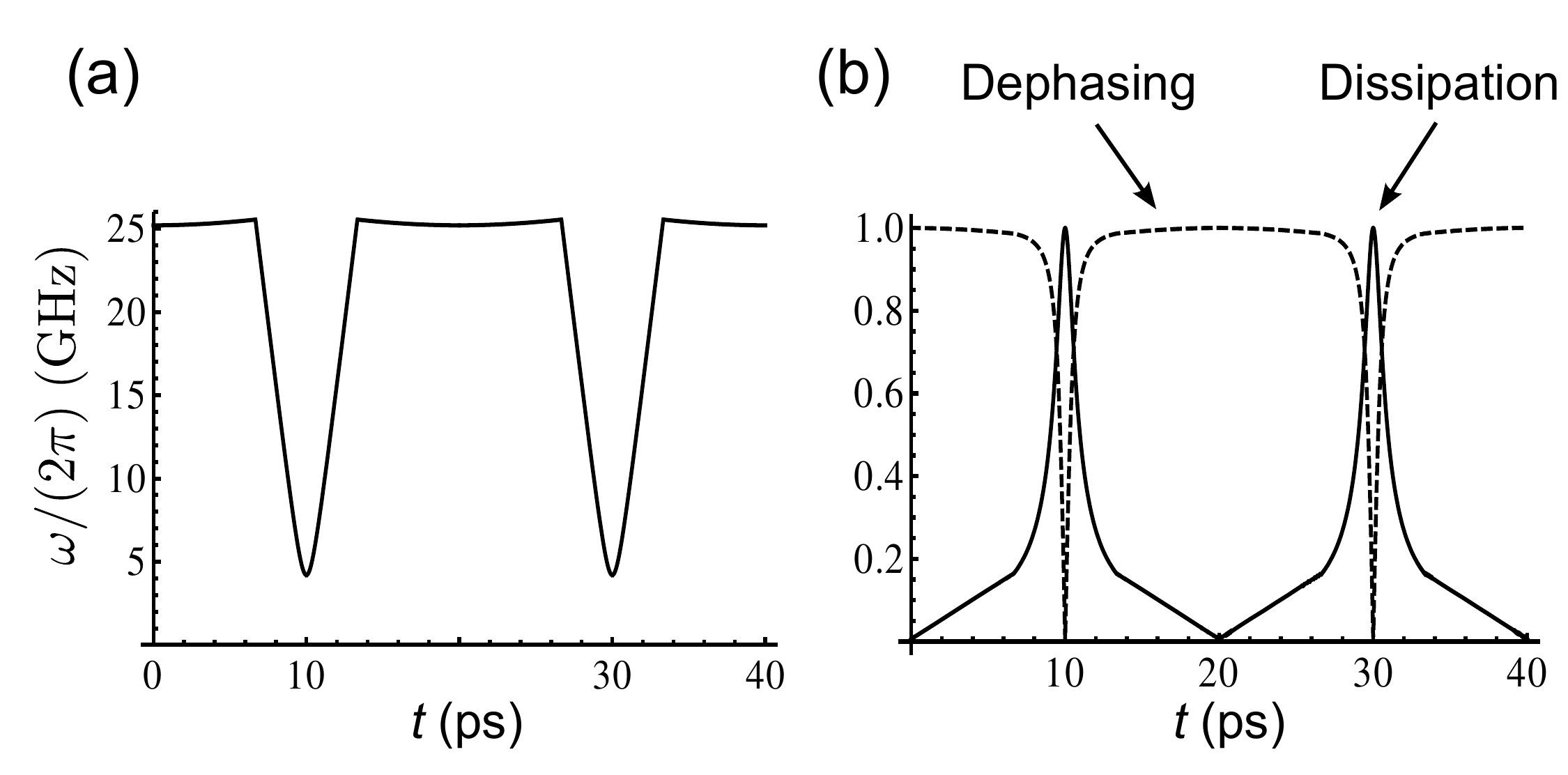} \vspace{-5mm}
		\caption{Panel (a) shows the energy splitting in the two--level approximation in units of $2\pi\hbar$. Panel (b) shows $|\!\bra{\phi_+}\sigma_z\ket{\phi_-}\!|$ (solid) and $\frac12 |\bra{\phi_+}\sigma_z\ket{\phi_+}-\bra{\phi_-}\sigma_z\ket{\phi_-}|$ (dashed), which relate to the dissipation and dephasing rate, respectively. \label{fig3}}
	\end{figure}

	In this system the main source of decoherence are fluctuations of the gate charge $n_g$, which according to \eqref{22} results in a coupling to the environment described by the operator $E_C\sigma_z$. Furthermore, we assume that the antisymmetrized environmental correlation spectrum $\frac12[\gamma(\omega)-\gamma(-\omega)]=\gamma_0\omega$ increases linearly with frequency, as is typically the case for voltage fluctuations. We also assume an environment at thermal equilibrium $\gamma(-\omega)=e^{-\hbar\omega/(k_BT)}\gamma(\omega)$, which  leads to $\gamma(\omega)=\gamma_0\omega\left/\left[ 1-\exp\!\left(\frac{-\hbar\omega}{k_BT}\right) \right]\right.$ as well as $\gamma(0)=\gamma_0k_BT/\hbar$, although dephasing is not important in a Cooper pair pump because the system starts in the ground state.

\begin{figure}[t]
		\includegraphics[width=0.9\linewidth]{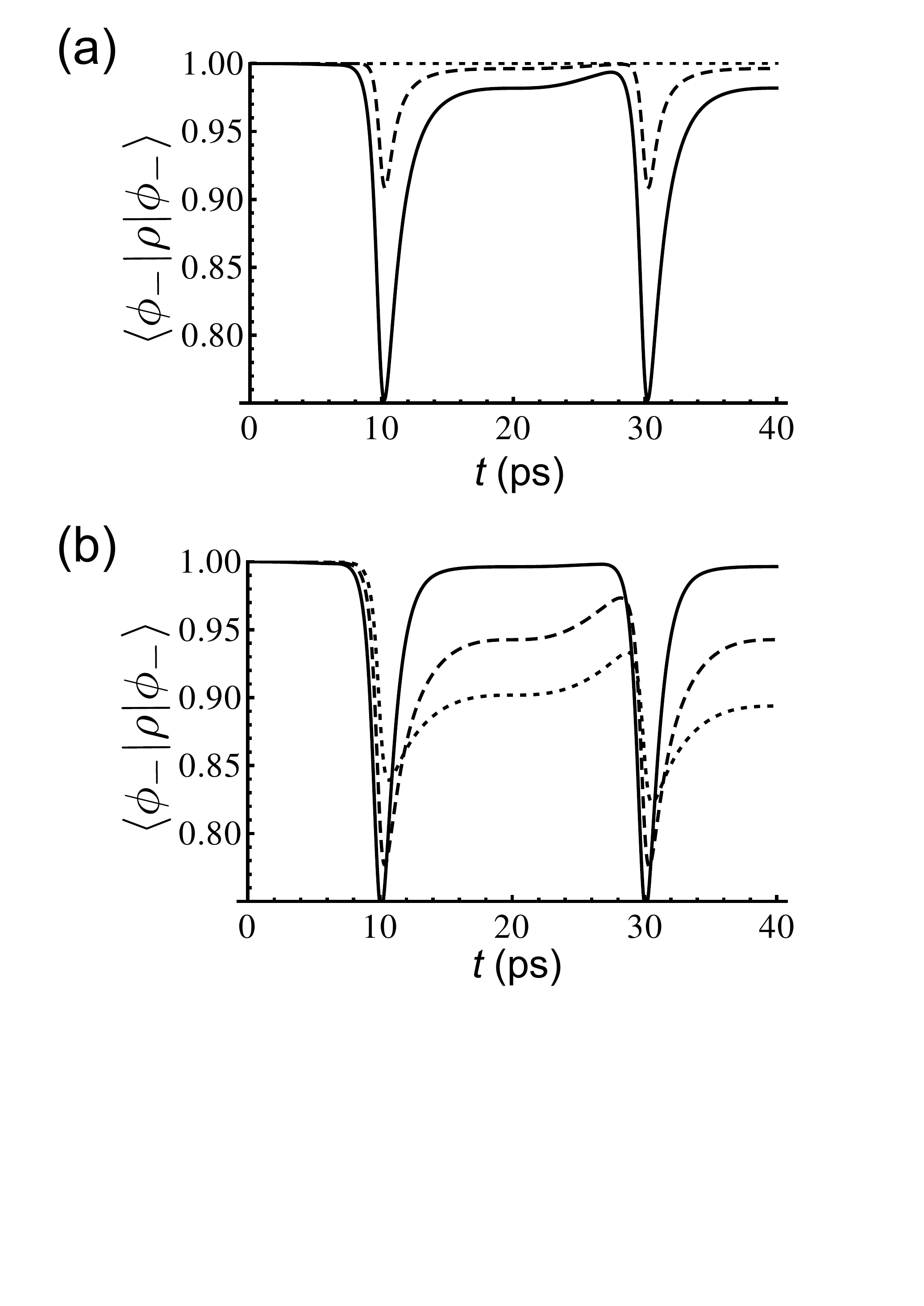}\vspace{-28mm}
		\caption{Population of the Floquet ground state during one pumping cycle plotted as a function of time in pico seconds. Initially the system is prepared in the Floquet ground state. In panel (a) we compare different temperatures $k_BT/(2\pi\hbar) =4$ (solid), 2 (dashed), 0 (dotted)~GHz at coupling to dissipative bathe (voltage fluctuations) of strength $E_C^2\gamma_0=0.1$, while in panel (b) we compare different coupling strengths $E_C^2\gamma_0=0.2$ (solid), $0.05$ (dashed), $0.02$ (dotted) at temperature $k_BT/(2\pi\hbar) = 4$~GHz.  \label{fig4}}
	\end{figure}

	It was already shown in~\cite{pek} that a zero temperature environment has no effect on ground state pumping. Here, we consider a finite temperature environment, which is able to excite the system. In Fig.~\ref{fig3}~(a), the level splitting of the Cooper pair sluice is plotted as a function of time for one pumping cycle. It is important to note that at times when the level splitting is small, i.e.\ when the system can be excited by a relatively low temperature environment, then the dissipative coupling to the gate voltage fluctuations has a pronounced maximum, as is seen in panel~(b) of Fig.~\ref{fig4}. Therefore, we can expect that voltage fluctuations with temperature $T=200~\rm{mK}\approx4~$GHz~$\times 2\pi\hbar/k_B$ already have a significant effect on the state of the system.
	
\paragraph*{\bf Transient phenomena.}

	The first few pumping cycles are not of importance in Cooper--pair pumping, as the experimentally measurable quantity is the averaged current after many cycles. However, in many other applications such as adiabatic quantum gates, one is interested mainly in the behavior of the system within one period. Therefore, we discuss here some interesting effects which happen before the system approached the quasi stationary state. Furthermore, this discussion might also be helpful in understanding the quasi stationary behavior studied below.

	In panel (a) of Fig.~\ref{fig4} we consider the system initially in the Floquet ground state and we plot the population of the Floquet ground state as a function of time for a fairly strong coupling to the dissipative bath for different temperatures. As predicted in~\cite{pek}, the system does not get excited if the environment has zero temperature (dotted line). As expected, the higher the temperature, the more populated the Floquet excited state becomes, in particular at times when the level splitting is small.

	In panel (b) of Fig.~\ref{fig4}, we plot the population of the Floquet ground state as a function of time for a temperature of about 200~mK and various environmental coupling strengths. If the coupling is strong (solid line), the system follows the instantaneous thermal mixture of the Floquet ground and excited states. For weaker couplings (dashed and dotted lines), the system gets less excited during the short time where the level splitting is small. Furthermore, the short time interval with dissipative coupling [see Fig~\ref{fig3}~(b)] is not sufficient for the system to return to the ground state once the level splitting is large.

	\begin{figure}[t]
		\includegraphics[width=0.8\linewidth]{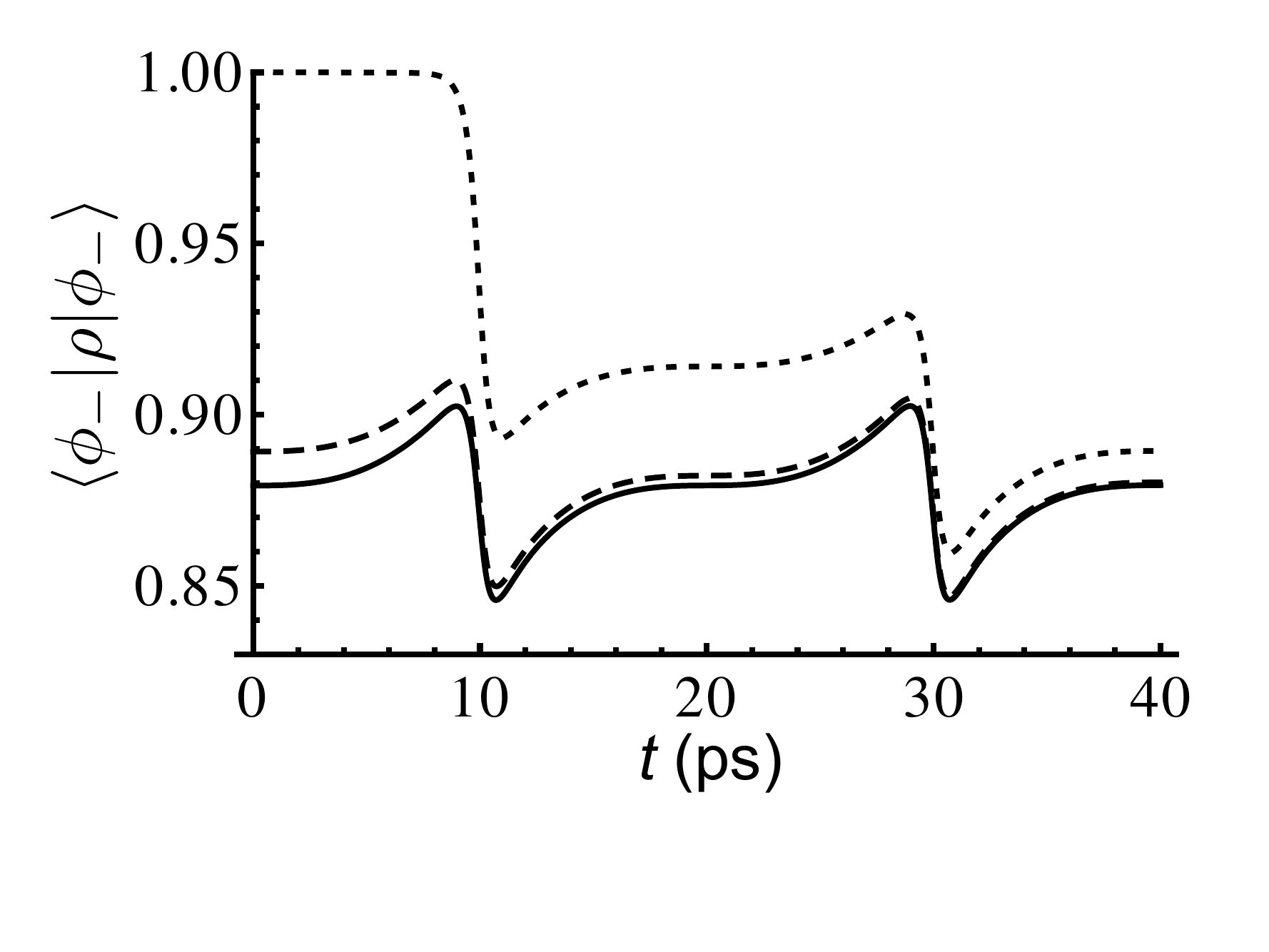}\vspace{-10mm}
		\caption{Population of the Floquet ground state for the first (dotted) and second (dashed) pumping cycles, as well as for the quasi stationary state (solid). We use the parameters $k_BT/(2\pi\hbar) =4$~GHz and $E_C^2\gamma_0=0.01$.   \label{fig5}}
	\end{figure}
	Finally, in Fig.~\ref{fig5} we use even weaker coupling to the environment such that the system needs several pumping cycles to approach a quasi stationary state. As mentioned in section~\ref{sec3}, this quasi stationary state still has time dependent populations of the two Floquet states.

\paragraph*{\bf Quasi stationary state.}

	\begin{figure}[t]
		\includegraphics[width=0.9\linewidth]{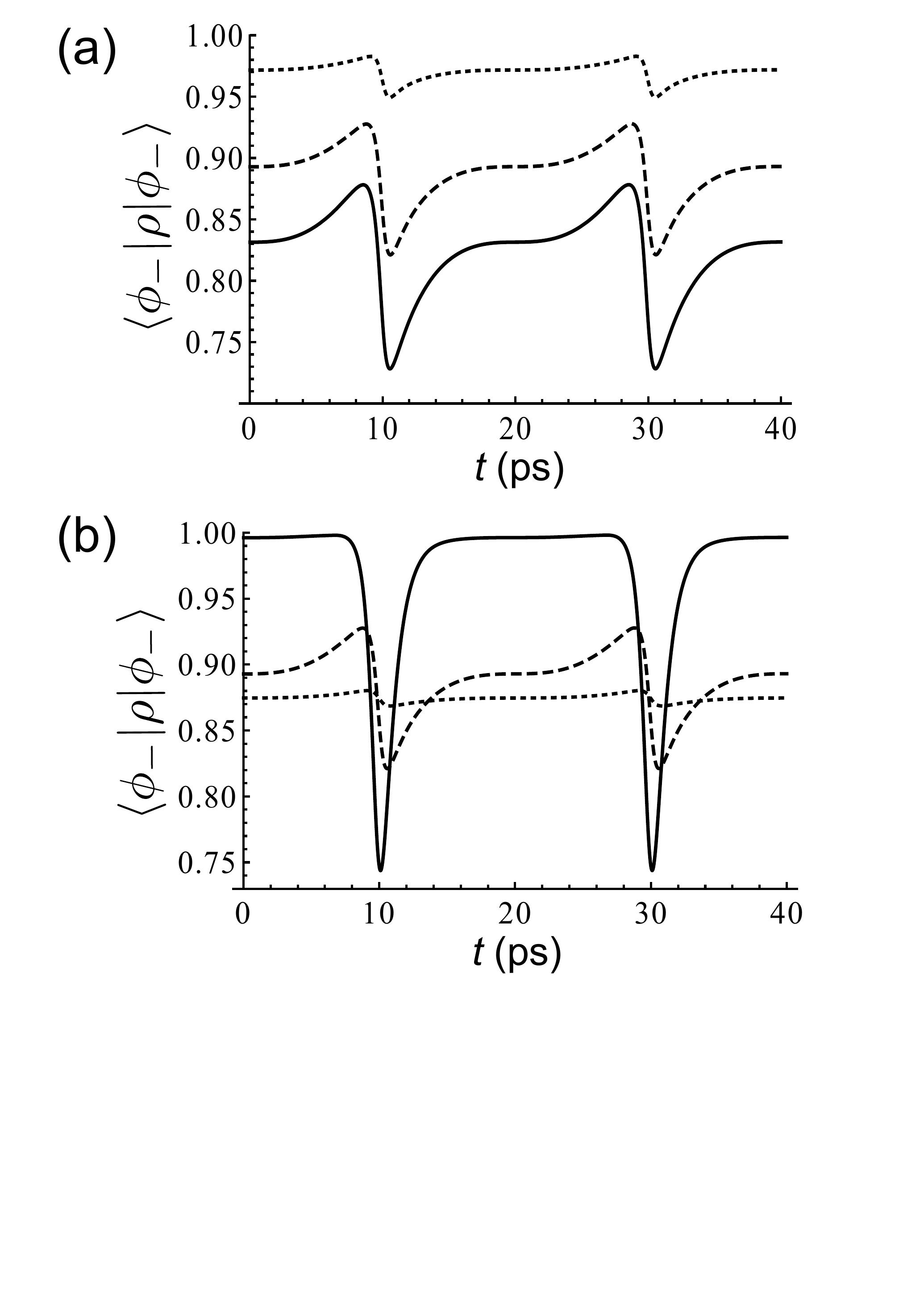}\vspace{-28mm}
		\caption{Population of the Floquet ground state in the quasi stationary state. In panel (a) we compare different temperatures $k_BT/(2\pi\hbar) =2$ (dotted), 4 (dashed), 6 (solid)~GHz for a coupling to dissipative bath (voltage fluctuations) with $E_C^2\gamma_0=0.1$. In panel (b) we compare different coupling strengths $E_C^2\gamma_0=0.2$ (solid), $0.02$ (dashed), $0.002$ (dotted) at temperature $k_BT/(2\pi\hbar) = 4$~GHz.  \label{fig6}}
	\end{figure}
	Now we assume that the system has already reached its quasi stationary state. In panel~(a) of Fig.~\ref{fig6}, we show the time dependence of the population of the Floquet ground state for various temperatures. Again, at times of a small level splitting the population of the Floquet excited state is increased, while at times of a large level splitting the system returns towards the Floquet ground state.

	In panel~(b), we compare the cases of different coupling strengths to the dissipative bath. If the coupling is strong (solid) the system closely follows a thermal mixture of the two Floquet states according to the instantaneous level splitting. On the contrary, if coupling to the environment is weak (dotted), it barely influence the system's state within one pumping cycle, which leads to a fairly constant population of each Floquet state. This is the limit which is described in Ref.~\cite{rus}. The dashed line represents the intermediate situation.

\paragraph*{\bf Current and pumped charge.}

	\begin{figure}[t]
		\includegraphics[width=\linewidth]{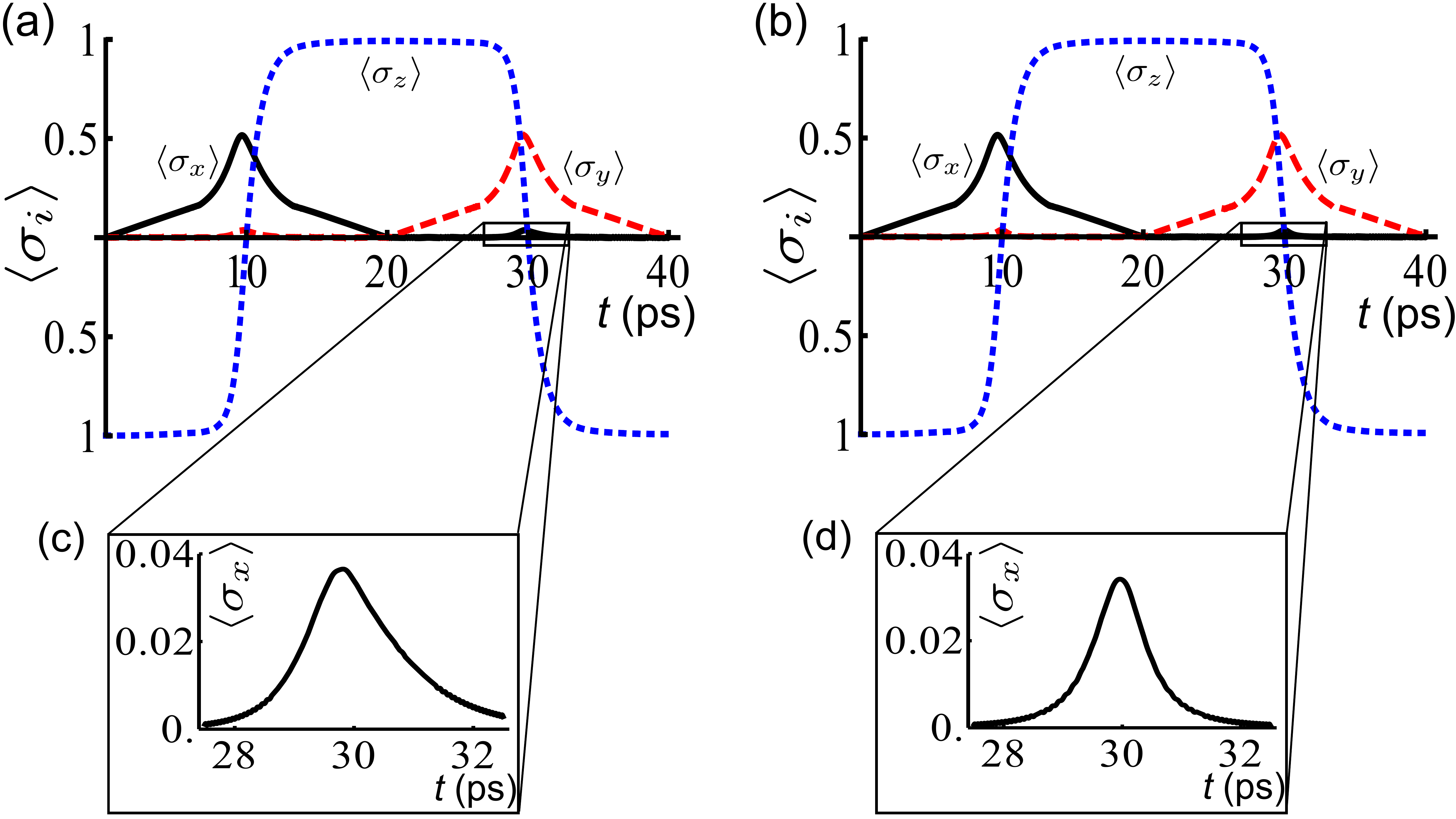}\vspace{-1mm}
		\caption{(Color online) The expectation values of the observables $\sigma_x$ (black, solid), $\sigma_y$ (red, dashed), and $\sigma_z$ (blue, dotted) are plotted as a function of time. For panel~(a) the density operator is calculated without using the secular approximation, while for panel (b) the secular approximation was applied. The similarity of the two panels verifies the accuracy of the secular approximation when being interested in the state of the system. The small peak of $\sigma_x$ at around $t=30$~ps is responsible for the current and is shown in detail in panels (c) and (d). We see that this peak is altered by the secular approximation, which explains why this approximation should be avoided when calculating the current. We use $E_C^2\gamma_0=0.2$ and $k_BT/(2\pi\hbar) =4$~GHz. For clarity, we use $J_{\rm{min}}=0$ to avoid supercurrent and to expect one transferred Cooper pair per cycle (see main text). Indeed, panel~(a) predicts 1.00 Cooper pairs, while performing the secular approximation [panel~(b)] leads to the incorrect value of 0.62 transferred Cooper pairs.
		\label{figError}}
	\end{figure}
	The operator of the current through the right Josephson junction is given by (see, e.g., Ref.~\cite{nis,mot2}) 
	\begin{eqnarray}
		I &=& -\frac{2e}{\hbar}\frac{\partial H} {\partial\varphi} \;=\; \frac{ieJ_R}{\hbar}
		\left(\begin{array}{cc}
			0 & -e^{-i\varphi} \\
			e^{i\varphi} & 0
		\end{array}\right)\ .
	\end{eqnarray}
	For $\varphi = \pi/2$ we obtain $I= -\frac{eJ_R}{\hbar}\sigma_x$.
	The expectation value of the current is rather small if the state of the system is close to the ground state and therefore small errors in the density matrix can lead to significant relative errors in the current. In particular, it was argued in~\cite{pek,sol} and discussed in detail in~\cite{mot3} that the secular approximation leads to incorrect values for the current and therefore to the non-conservation of charge. In Fig.~\ref{figError} we analyze the validity of the secular approximation. We observe that the secular approximation is very good for the expectation values of all three observables, $\sigma_x$, $\sigma_y$, and $\sigma_z$. Thus the quantum state of the system is well approximated. However, the small deviations in $\langle \sigma_x\rangle$ shown in Fig.~\ref{figError} are sufficient to significantly modify the result for the pumped charge, when integrated over long time. 

	We discuss the situation shown in Fig.~\ref{figError} in more detail. To facilitate the qualitative understanding we assume, first, a vanishing residual Josephson coupling $J_{\rm{min}}=0$ and a coupling to the environment which is sufficiently strong for the state to closely follow the instantaneous thermal equilibrium. The temperature is of the order of the minimal energy gap, but much lower than the energy gap at the beginning and the end of each half--cycle [see panel~(a) of Fig.~\ref{fig3}]. At the beginning of the cycle the system will therefore be in the ground state with no excess Cooper pair on the island. When the system moves through the first avoided level crossing, the energy gap is small enough for the excited state to be populated. Nevertheless, after the first half--cycle is completed, the energy splitting is large again and the system will again be in the ground state, which now has one excess Cooper pair on the island. Because $J_R=0$ during the the first half--cycle, this Cooper pair came from the left lead. With the same reasoning, the Cooper pair leaves the island to the right lead during the second half cycle. Therefore, in this parameter regime we should obtain exactly one transferred Cooper pair per pumping cycle, much like in the ground state pumping without coupling to an environment. Indeed, avoiding the secular approximation we obtain exactly one Cooper pair per cycle [panel~(a) of Fig.~\ref{figError}], while performing the secular approximation [panel~(b)] leads to the incorrect value of 0.62 transferred Cooper pairs.

	This failure of the secular approximation discussed above is not necessarily connected to the time dependence of the Hamiltonian. To illustrate this, consider a time independent Hamiltonian with $J_L=0$ and $J_R\neq0$ and assume the system to be relaxing from some initial non-equilibrium state to the thermal equilibrium state. In this process the populations of the excited state and the ground state change in time. Since these states have different expectation values of charge on the island, some current must flow. However, for $J_L=0$ only off-diagonal elements of the density operator in the energy basis are responsible for current and in the secular approximation the off-diagonal elements are never generated if they were not present in the initial state.
This explains why the secular approximation must fail to take into account contributions to the current which are induced by dissipative effects.

To guarantee the correctness of the transferred charge, in the following we use the master equation \eqref{14A} which was derived without the use of the secular approximation.

	In absence of dissipation the pumped charge $Q_p$ and the charge transferred due to the 
	super current $Q_s$ were found~\cite{mot,pek2}
 to be given by
		\begin{eqnarray}
		Q_p &=& -2e\left(1-2\frac{J_{\rm{min}}}{J_{\rm{max}}}\cos\varphi\right) \\
		Q_s &\propto& \mathcal T I_{\rm min}\sin\varphi,
	\end{eqnarray}
	where $I_{\rm min} \equiv 2e J_{\rm min}/\hbar$ is the residual critical current of the system. 
	In the strictly adiabatic limit, these formulas need not to be modified if the environment has zero temperature~\cite{pek}. Our choice of $\varphi=\pi/2$ then leads to exactly one pumped Cooper pair per cycle, and any deviation from that ideal value is because the finite temperature environment is able to excite the Cooper pair sluice.

	\begin{figure}[t]
		\includegraphics[width=0.8\linewidth]{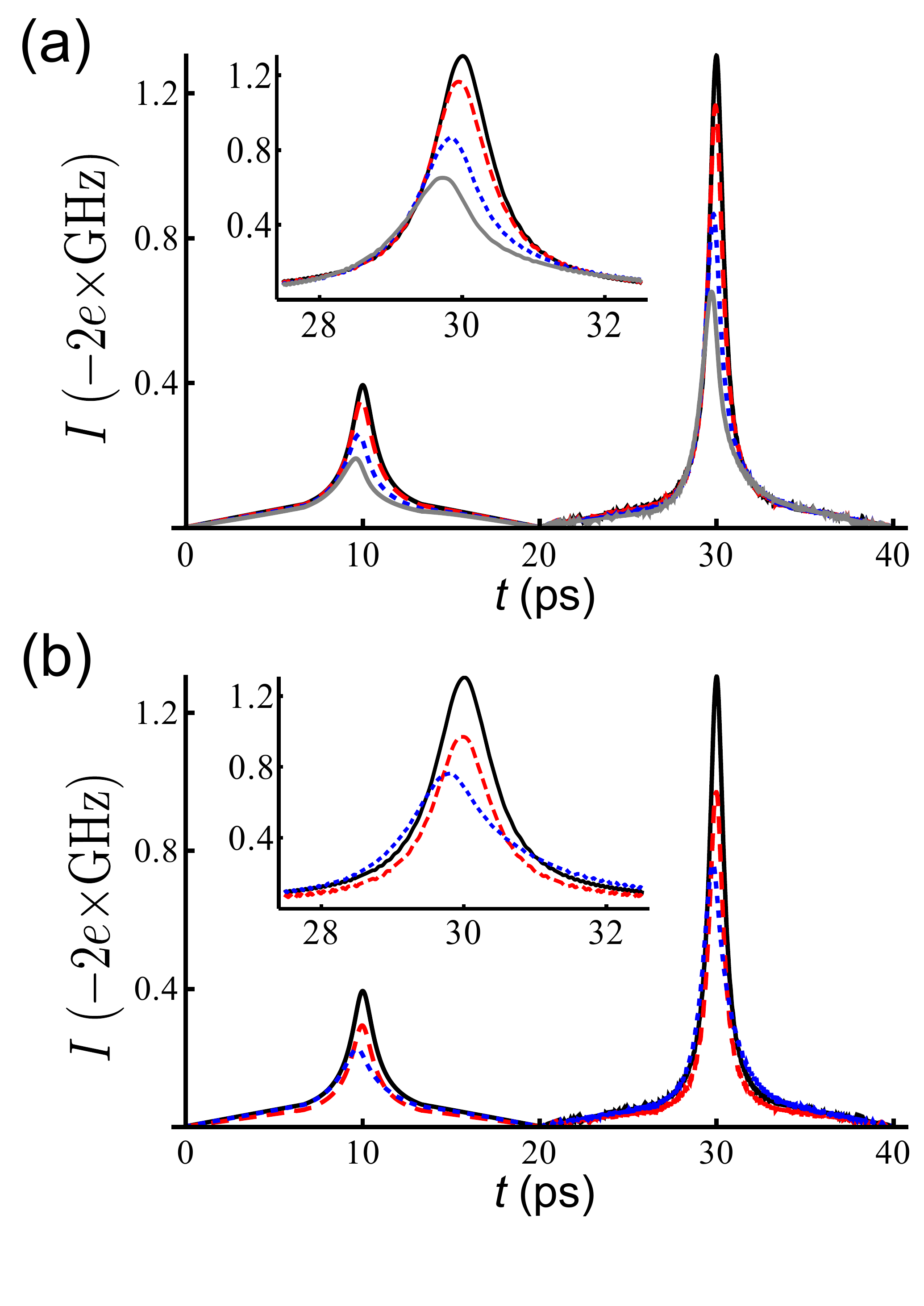}\vspace{-8mm}
		\caption{(Color online) The total current is plotted as a function of time. Panel (a) compares different temperatures $k_BT/(2\pi\hbar) =0$ (black, solid), 2 (red, dashed), 4 (blue, dotted), 6 (gray, solid)~GHz at fixed coupling strength $E_C^2\gamma_0=0.05$. Panel (b) compares different coupling strength $E_C^2\gamma_0=0$ (black), 0.005 (red), 0.1 (blue) at fixed temperature $k_BT/(2\pi\hbar) =4$~GHz. The insets highlight the second current peak. \label{fig7}}
	\end{figure}
\begin{figure}[t]
		\includegraphics[width=\linewidth]{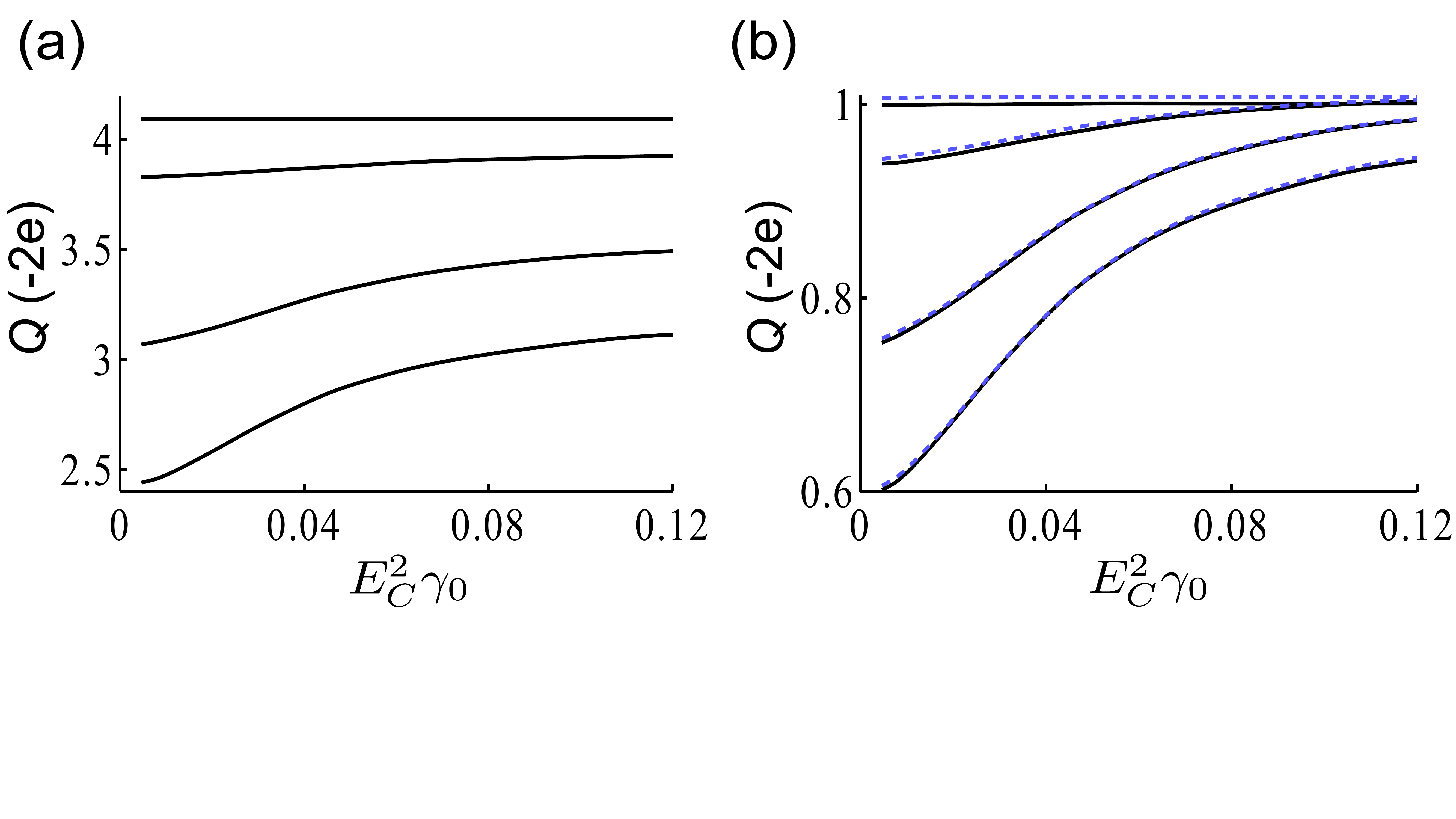}\vspace{-13mm}
		\caption{(Color online) Panel~(a) shows the total transferred charge per pumping cycle in units of Cooper pairs as a function of the coupling strength $E_C^2\gamma_0$ to the environment for the temperatures $k_BT/(2\pi\hbar) =$0~GHz, 2~GHz, 4~GHz, 6~GHz (from top to bottom). The residual coupling is $J_{\rm{min}}=0.03J_{\rm{max}}$. Panel~(b) shows the pumped charge per cycle (solid curves), which is calculated from the difference of the total charges obtained by traversing the pumping cycle in opposite directions. For comparison (blue, dashed curves) we also show the transferred charge per cycle without residual current $J_{\rm{min}}=0$, which shows that the pumped charge is barely influenced by the residual coupling $J_{\rm{min}}$. \label{fig8}}
	\end{figure}

	In Fig.~\ref{fig7} we plot the current in units of Cooper pairs per picosecond as a function of time, assuming that the system has already undergone enough cycles to be in the quasi--stationary state. In panel~(a) we compare different temperatures and we find that as the temperature increases, the current through the sluice decreases. The reason is, that the Floquet--excited state carries supercurrent in the opposite direction to that of the ground state. In panel~(b), we compare different coupling strengths to the dissipative bath (voltage fluctuations), and similarly find that the peak current decreases with the coupling strength. However, the peak width increases with coupling strength such that the total transferred charge per cycle increases with the coupling strength (see panel~(a) of Fig.~\ref{fig8}). This behavior can be explained qualitatively by looking at Fig.~\ref{fig6}~(b). At the time of maximal current, the population of the ground state decreases with growing coupling strength, therefore leading to a decreased peak current. But at other times, the population of the ground state increases with growing coupling strength, as does therefore the current, resulting in wider current peaks.

	Panel (a) of Fig.~\ref{fig8} shows the total transferred charge per pumping cycle $Q_{\rm{tot}}=Q_s+Q_p$ plotted as a function of the coupling strength to the dissipative bath for different temperatures. Again, the charge decreases with temperature, but it slightly increases with $E^2_C\gamma_0$. Panel~(b) shows the corresponding pumped charge $Q_p$, which is obtained from $Q_p=(Q_{\rm{tot}}-\overline Q_{\rm{tot}})/2$, where $\overline Q_{\rm{tot}}$ is the transferred charge if the pumping cycle is traversed in the opposite direction which changes the sign of the pumped charge but leaves the supercurrent unchanged. As expected, the pumped charge decreases with temperature. The dependence on the coupling strength to the environment shows an increase with the growing coupling. This supports the conclusion of~\cite{pek} that a strong coupling to the environment might be helpful for Cooper pair pumping, and can be explained by the fact that at times when the level splitting is large, the system relaxes more effectively to the ground state. As elaborated before for $J_{\rm{min}}=0$, for one pumped Cooper pair per cycle, it is sufficient if the system is in the ground state at the beginning of each half cycle. We further plotted the transferred charge for $J_{\rm{min}}=0$ which shows that the pumped charge is barely influenced by $J_{\rm{min}}$. For $k_BT/(2\pi\hbar) =$6~GHz and strong coupling the pumped charge does not converge to one Cooper pair, but approaches a somewhat lower value. That is because a higher temperature environment is able to excite the system even at the end of each half cycle when the energy splitting is large.

\section{Conclusions}

	In conclusion, we have explicitly taken into account the time dependence of an adiabatically and cyclically evolving Hamiltonian to develope a master equation which is valid if the environment is Markovian. We found that decoherence takes place in the time dependent Floquet basis and that the decoherence rates depend on time through the instantaneous level splittings as well as the instantaneous coupling to the noise.

	We first applied our theory to a spin in an adiabatically changing magnetic field, where we found that the spin does not relax into the instantaneous ground state, but the slightly different Floquet ground state which is very close to the superadiabatic ground state. We then tested our theory on the example of the Cooper pair sluice, where the time 
dependence of the decoherence rates is very pronounced. This results in new features at finite temperatures and modifies the pumped charge which is measurable in current experiments~\cite{mot2,mot}. We further explain why the current has to be calculated without using the secular approximation despite the fact that this approximation is very well justified if one is interested only in the state (density matrix) of the system.

\acknowledgments{The authors are grateful to S. Pugnetti, P. Solinas, M. M\"ott\"onen, and J. Pekola for valuable discussions. This work was funded by the EU FP7 GEOMDISS project.}


\begin{thebibliography}{99}
\bibitem{notation} There is no consistent terminology in the literature regarding terms like decoherence. We use the term decoherence to describe all effects induced by an environment. This includes the decrease of the off--diagonals of the density matrix in the energy basis (dephasing) as well as the relaxation of the diagonals towards a steady state (dissipation, energy relaxation).
\bibitem{footnote} The pulse shape is not optimized for better adiabaticity, but we use this shape to be able to 
compare with recent results in the literature.
\end{thebibliography}
\end{document}